\pdfoutput=1
\documentclass[a4paper,aps,pdftex,floatfix,citeautoscript,groupedaddress,pra,showpacs,twocolumn]{revtex4-1}

\usepackage{amsfonts,amsmath,amssymb}
\usepackage{graphicx}
\usepackage{hypernat}
\usepackage[utf8]{inputenc}
\usepackage{textcomp}
\usepackage{xspace}
\usepackage[hyperfootnotes=true]{hyperref}
\usepackage{placeins}
\usepackage{todonotes}
\usepackage{overpic}
\hypersetup{colorlinks,citecolor=green,urlcolor=blue}

\newlength{\figwidth}
\newcommand{\ie}{i.\,e.\ }%

\newcommand{\costhetasqtd}{\ensuremath{\langle\cos^2\theta_{2D}\rangle}}
\newcommand{\costhetasq}{\ensuremath{\langle\cos^2\theta\rangle}}
\newcommand{\ket}[1]{\ensuremath{\left|\mathrm{#1}\right>}}

\begin{document}

\title{Stark-selected beam of ground-state OCS molecules \\ characterized by revivals of
   impulsive alignment}%
\author{Jens H. Nielsen$^1$}%
\author{Paw Simesen$^1$}%
\author{Christer Z. Bisgaard$^2$}%
\author{Henrik Stapelfeldt$^{3,4}$}%
\email{henriks@chem.au.dk}%
\affiliation{%
   $^1$Department of Physics and Astronomy, University of Aarhus, 8000 Aarhus C, Denmark \\
   $^2$Department of Photonics Engineering, Technical University of Denmark, 2800 Kgs.\ Lyngby, Denmark \\
   $^3$Department of Chemistry, University of Aarhus, 8000 Aarhus C, Denmark \\
   $^4$Interdisciplinary Nanoscience Center (iNANO), University of Aarhus, 8000 Aarhus C, Denmark}%

\author{Frank Filsinger$^5$}%
\author{Bretislav Friedrich$^5$}%
\author{Gerard Meijer$^5$}%
\author{Jochen K\"upper$^{5,6,7}$}%
\email{jochen.kuepper@cfel.de}%
\affiliation{%
   $^5$Fritz-Haber-Institut der Max-Planck-Gesellschaft, Faradayweg 4-6, 14195 Berlin, Germany \\
   $^6$Center for Free-Electron Laser Science, DESY, Notkestrasse 85, 22607 Hamburg, Germany \\
   $^7$University of Hamburg, Luruper Chaussee 149, 22761 Hamburg, Germany}%

\date{\today}%
\begin{abstract}\noindent%
   We make use of an inhomogeneous electrostatic dipole field to impart a quantum-state-dependent
   deflection to a pulsed beam of OCS molecules, and show that those molecules residing in the
   absolute ground state, $X ^1\Sigma^+$, $\ket{00^00}$, $J=0$, can be separated out by selecting
   the most deflected part of the molecular beam. Past the deflector, we irradiate the molecular
   beam by a linearly polarized pulsed nonresonant laser beam that impulsively aligns the OCS
   molecules. Their alignment, monitored via velocity-map imaging, is measured as a function of
   time, and the time dependence of the alignment is used to determine the quantum state composition
   of the beam. We find significant enhancements of the alignment (\costhetasqtd $= 0.84$) and of
   state purity ($> 92$\%) for a state-selected, deflected beam compared with an undeflected beam.
\end{abstract}
\maketitle

\noindent%
The ability to produce ensembles of atoms and molecules with a narrow distribution of quantum states
has been a game-changer in atomic, molecular and optical physics, past and present. Recent examples
from molecular physics include crossed beam
scattering~\cite{kopin,termeulen,brouard,Scharfenberg-Meijer} and photodissociation dynamics studies
at high resolution \cite{rakitzis_directional_2004}, as well as the work done with and on cold and
ultracold molecules \cite{cold-molecules-book:2009,DeMille_2009}. The techniques developed to
produce molecules in (nearly) single quantum states include multipole focusing
\cite{bennewitz_fokussierung_1955, Stolte:BBGPC86:413, Reuss:StateSelection,
   parkerbernstein:1989:arpc, Putzke:PCCP:submitted}, Stark
deceleration~\cite{bethlem:1999:prl,van_de_meerakker_taming_2008}, Zeeman
deceleration~\cite{Merkt-PRA-2007, Narevicius-PRA-2008} -- all of which isolate molecules that are
initially populated in a particular quantum state -- and buffer gas cooling \cite{Doyle-Nature-1998,
   Motsch-Rempe-PRL-2009} -- which lowers the temperature of a sample such that the resulting
thermal state-distribution contains essentially only the absolute ground state. Alternatively,
ultracold alkali atoms can be photo- or magneto-associated and form ultracold homo- or heteronuclear
diatomic molecules occupying a single electronic, vibrational and rotational state
\cite{Naegerl-Science-2008,Denschlag:PRL:2008} including, in certain cases, the absolute ground
state \cite{Wester:Weidemuller:PRL:2008,Ye:Science:2008}. All methods offer unique opportunities but
are also subjected to limitations in terms of the type of molecules that the methods apply to or the
particular quantum states that can be selected.

Here we demonstrate an alternative method, with a history reaching back to the 1920s
\cite{kallmann_ber_1921,wrede_ber_1927,stern_zur_1926}, to produce molecules in a single quantum
state. Like multipole focusing and Stark or Zeeman deceleration it is based on selecting the
molecules that are initially residing in a specific quantum state. We employ a dipolar deflection
field to disperse a well-expanded, nearly monoenergetic beam of OCS molecules according to their
state-specific electric dipole moments. The inhomogeneous electric field inside the deflector has an
almost constant gradient over a large area surrounding the molecular beam axis and enables
dispersion of the rotational quantum states of OCS. In particular, we separate out the ground
rotational state, $J=0$, of the electronic and vibrational ground state, $X ^1\Sigma^+$ and
$\ket{00^00}$, and thereby produce a molecular beam of OCS($X ^1\Sigma^+, \ket{00^00}, J=0$) with a
purity in excess of 92\%. Since $^{16}$O$^{12}$C$^{32}$S has zero nuclear spin, the ground state
obtained is free of hyperfine structure. The purity of the beam was characterized by observing the
time dependence of the nonadiabatic alignment
\cite{ortigoso-friedrich:1999,seideman_revival_1999,rosca-pruna:2001:prl,stapelfeldt_colloquium:_2003}
produced by the interaction of a linearly polarized nonresonant pulsed laser beam with the
anisotropic OCS molecules.

In what follows, we first present an outline of the experimental setup. Then we describe the
dispersion of the molecular beam achieved by the dipole deflector. The nonadiabatic alignment of the
molecules and its use to characterize state purity of the molecular beam is described next. Finally,
we draw conclusions from our work.

\begin{figure}
  \includegraphics[width=\figwidth]{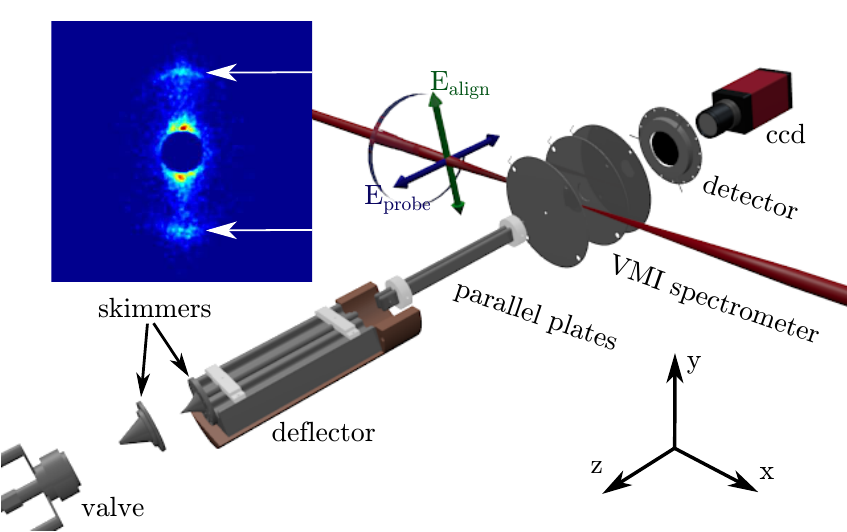}
  \caption{Schematic of the experimental setup. The inset shows an S$^+$ ion image recorded when the
     molecules are aligned along the y-axis (t = 40.6 ps). The arrows indicate images of the
     $\textrm{S}^{+}$ + $\textrm{CO}^{+}$ Coulomb-explosion channel employed to determine the degree
     of alignment.}
  \label{fig:expsetup}
\end{figure}
The experimental setup, detailed earlier
\cite{filsinger_quantum-state_2009,holmegaard_laser-induced_2009}, is shown in
\autoref{fig:expsetup}. A pulsed molecular beam is produced by expanding a mixture of 10~bar of Neon
and 1~mbar of OCS into vacuum through a 250 $\mu$m diameter nozzle in a pulsed valve. The beam is
collimated by a skimmer and sent into a 15 cm-long electrostatic deflector, which exerts a force on
the molecules along the y-axis, cf. \autoref{fig:expsetup}. After the exit of the deflector the
molecules travel another 17~cm through a region between two parallel electrostatic plates (field
strength: 2 kV/cm) to preserve a field-quantization axis before they enter into a velocity map
imaging (VMI) spectrometer. Here the molecular beam is crossed by the pulsed alignment and probe
laser beams propagating along the x axis. The laser beams, operating at a (nonresonant) wavelength
of 800 nm, are focused on the molecular beam with waists ($\omega_0$) of 35 and 25 $\mu$m,
respectively. The alignment pulses, linearly polarized along the y-axis, have a duration of 330 fs.
This is shorter than the rotational period, $\tau_{\text{rot}}=82$ ps, of OCS($X ^1\Sigma^+,
\ket{00^00}$) by about a factor of 250, ensuring that the anisotropic polarizability interaction
with the molecule is nonadiabatic (in fact well within the sudden regime). The probe pulses,
linearly polarized along the z-axis, are 30 fs long and ionize the molecules via nonresonant
multiphoton absorption. The positive ions thus produced are accelerated toward an imaging detector;
the 2D ion images constitute the primary data obtained in the experiment.

\begin{figure}
  \begin{overpic}[width=\figwidth]{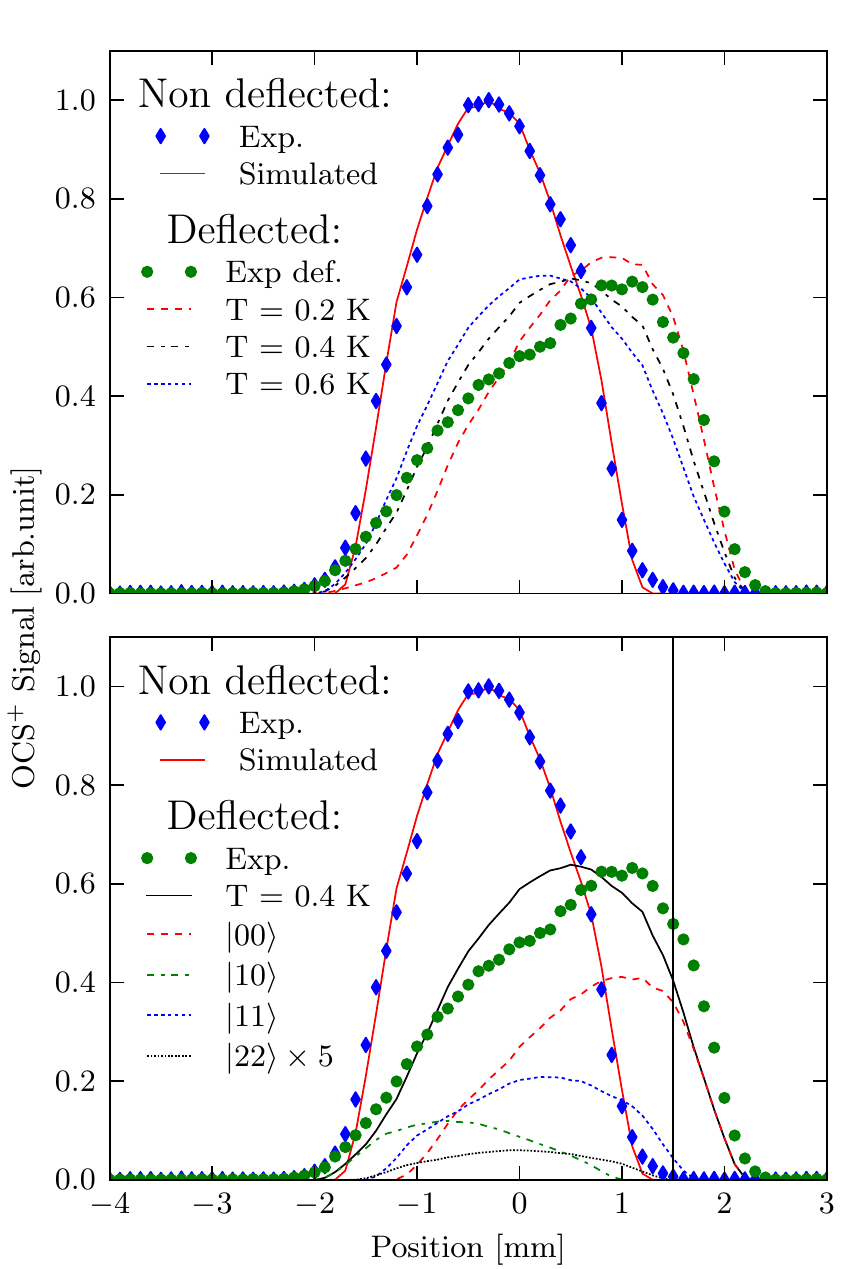}
    \put(60,93){(a)}
    \put(60,46.8){(b)}
  \end{overpic}
  \caption{Molecular beam intensity profiles along the y-axis. (a) Experimental data (points) and
     simulated profiles (curves) for different rotational temperatures. (b) Experimental data
     (points) and simulated profiles (curves) at $T_\text{rot}= 0.4$ K for individual rotational
     states. The profile for the $\ket{22}$ state is scaled up by a factor of 5. Vertical line
     indicates the position of the laser foci in the alignment experiment.}
  \label{fig:deflection}
\end{figure}
The effect of the deflector on the molecular beam was characterized by measuring the intensity
profile of the molecular beam along the (vertical) $y$ axis, see \autoref{fig:deflection}. This was
done by recording the $\textrm{OCS}^+$ signal, arising from the ionization of OCS by the probe pulse
as a function of the vertical position of the laser focus, see \autoref{fig:expsetup}. When the
deflector is turned off, the vertical width of the molecular beam is about 2.2 mm,
mainly due to the collimation of the beam by the skimmers, which precede the deflector. When the
deflector is turned on, the molecular beam profile broadens and shifts upwards, toward higher field
strengths (corresponding to higher values of $y$). The molecular beam profiles were modeled by a
Monte-Carlo simulation. For every relevant eigenstate of the initial OCS packet we generated a test
sample of initial values of molecules in phase-space ($x,y,z,v_\text{x},v_\text{y},v_\text{z}$ and
then performed (classical) trajectory simulations through the inhomogeneous electric field of the
deflector. This yields a vertical beam profile for the given molecular quantum state. These profiles
for individual states were averaged with a weight according to their populations -- according to a
Boltzmann distribution for different rotational temperatures $T$ of the beam and their degeneracy.
In \autoref{fig:deflection}\,(a) the resulting profiles for 0.2~K, 0.4~K, and 0.6~K are given along
with the experimental curve. A comparison of the measured and simulated data indicates that the
$T=0.4$~K curve represents the experimental data best. However, a non-negligible discrepancy
remains, particularly on the left tail of the deflection profile, corresponding to negative values
of $y$. Likely culprits are potentially increased populations of the
low-field seeking $\ket{10}$ and $\ket{20}$ states (with theoretical populations at $T= 0.4$ K of
$13\%$ and $1.2\%$, respectively). This is illustrated in \autoref{fig:deflection}(b), which shows
the Boltzmann-weighted profiles of the individual Stark states at $T=0.4$ K. An excess of the
$\ket{10}$, $\ket{20}$ states is consistent with the experimental observations made for the
impulsive alignment of OCS, described below. For the case of the temperature-independent direct
(non-deflected) beam, the simulations reproduce the experimental data accurately.

\autoref{fig:deflection}(b) further shows that of the upward-deflected high-field-seeking states,
the \ket{00} state deflects the most, followed by the \ket{11} state, and the \ket{22} state; the
low-field-seeking \ket{10} state deflects downwards. At $y = 1.5~\textrm{mm}$, $89~\%$ of the
molecules are in the \ket{00} state and $11~\%$ in the \ket{11} state, while the population of the
\ket{22} state is negligible.

In order to characterize the quantum state composition of the deflected molecular beam, we conducted
a nonadiabatic alignment experiment: At time $t = 0$, the molecular beam is irradiated by an
alignment pulse, see \autoref{fig:expsetup}. The peak intensity of
$9.6\times10^{12}~\textrm{W}/\textrm{cm}^2$ is low enough to preclude any detectable ionization of
OCS. Within its focal volume, the alignment laser pulse nonadiabatically excites each molecule to a
nonstationary, rotational wave packet, which undergoes periodic revivals, as does
the concomitant alignment. The rotational wave packet dynamics is probed by irradiating the
molecules with a probe pulse ($\textrm{I}_{\textrm{probe}} =
4.8\times10^{14}~\textrm{W}/\textrm{cm}^2$) at a time $t$. The probe pulse double-ionizes some of
the OCS molecules, triggering their Coulomb explosion into $\textrm{CO}^+$ and $\textrm{S}^+$ ion
pairs. This particular fragmentation channel can be identified by the recoil velocity as a pair of
radially displaced half-rings in the outermost region of the $\textrm{S}^+$ ion images, such as
those shown in the inset of \autoref{fig:expsetup}.

In keeping with our previous work, we assume that the Coulomb explosion occurs rapidly enough for
the axial recoil approximation to apply, in which case the emission direction of the ions is
straightforwardly related to the alignment of the molecule at the instant of ionization. We quantify
the alignment attained by the (experimental) degree of alignment, $\costhetasqtd$, where
$\theta_{2D}$ is the angle between the velocity vector of the $\textrm{S}^+$ ion in the detector
plane and the polarization plane of the alignment laser pulse.

\begin{figure}[ht]
  \includegraphics[width=\figwidth]{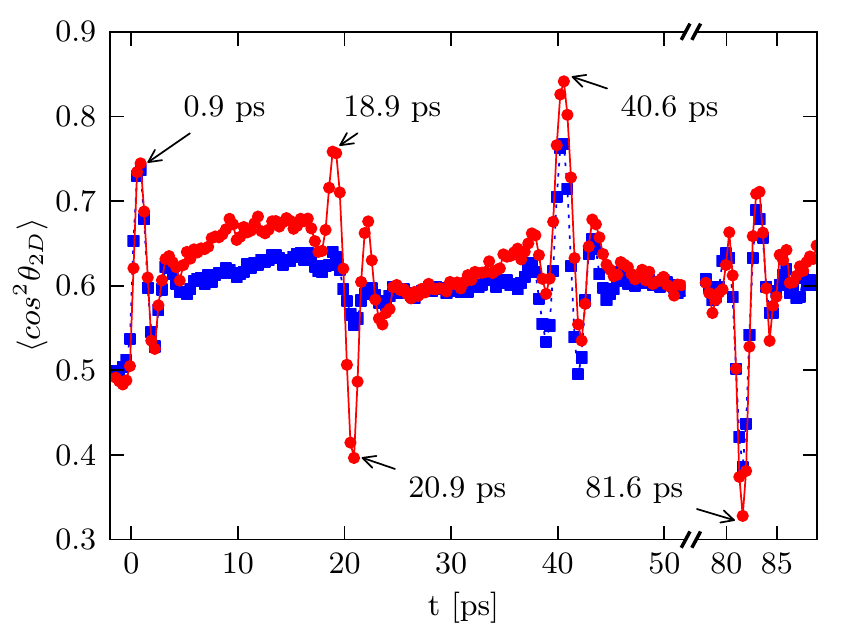}
   % \begin{overpic}{exp_aligment_break}
   %   \put(15,66){(a)}
   %   \put(90,66){(b)}
   % \end{overpic}
    \caption{Alignment dynamics of OCS molecules, represented by the time dependence of the
      alignment degree, $\langle\cos^2\theta_\text{2D}\rangle$. The origin, $t=0$, of the time scale
      is defined by the moment of arrival of the alignment pulse. Results obtained either with the
      direct beam or the deflected beam (at $y =1.5$ mm) are shown, respectively, by blue squares or
      red circles.}
  \label{fig:nonadexp}
\end{figure}
\autoref{fig:nonadexp} displays the rotational wave packet dynamics, represented by the dependence
of the alignment degree $\costhetasqtd$ on time over the intervals $- 2~\textrm{ps} < t <
52~\textrm{ps}$ and $77~\textrm{ps} < t < 89~\textrm{ps}$. The two blue trace pertains to
undeflected (deflector off) and the red trace to the deflected (deflector on) molecules for the
laser foci placed at $y = 1.5$~mm, cf.\ \autoref{fig:deflection}(b). The most prominent features of
the undeflected-beam trace are the prompt alignment arising shortly after the arrival (and passing)
of the alignment pulse near $t=0$, the half-period revival centered at $t = 40.6$ ps, and the first
full-period revival centered at $t = 81.6$ ps.

\begin{figure}
    \begin{overpic}[width=\figwidth]{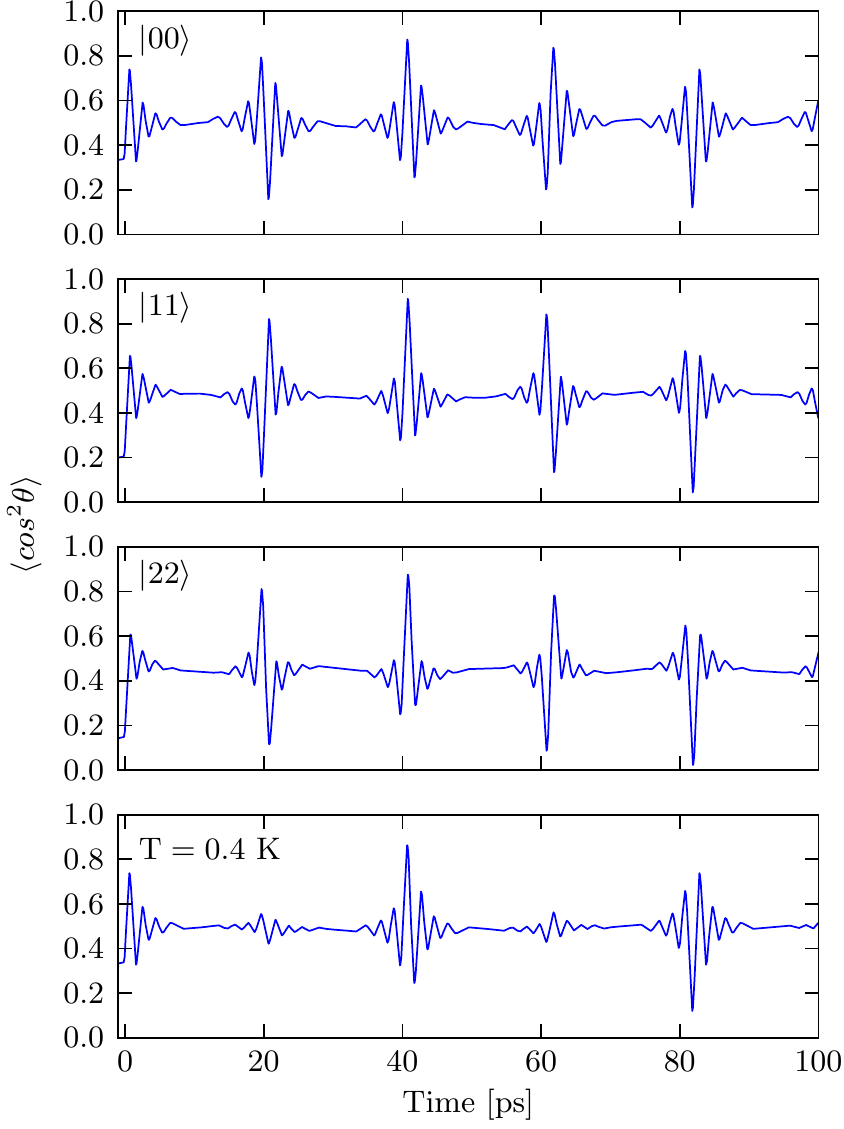}
      \put(67,95.5){(a)} \put(67,72){(b)} \put(67,48){(c)} \put(67,24.5){(d)}
    \end{overpic}
    \caption{Panels (a)-(c) show time dependence of the alignment cosine,
       $\langle\cos^2\theta\rangle_{JM}$, of OCS calculated over a rotational period of the molecule
       for different initial rotational states $|JM\rangle$. Panel (d) shows the time dependence of
       the ensemble-averaged alignment cosine $\langle\costhetasq\rangle$ of OCS at a rotational
       temperature $T=0.4$ K. Focal volume averaging was accounted for by including an
       $\textrm{I}^3$ probe detection efficiency based on measurements of the intensity dependence
       of the ionization yield.}
    \label{fig:nonadsim}
\end{figure}
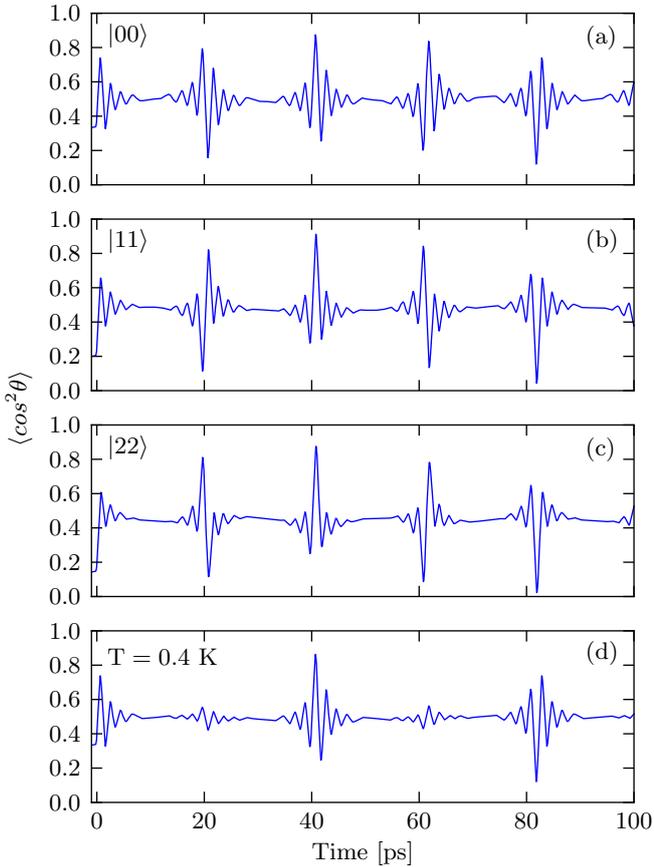

In order to account for the observed alignment dynamics, we solved numerically the time-dependent
Schr\"odinger equation for the interaction of a linearly polarized laser field with the anisotropic
polarizability of the OCS molecule \cite{ortigoso-friedrich:1999,bisgaard_alignment_2006}, and
evaluated the state-specific expectation values of the alignment cosine, $\costhetasq_{JM}$ for each
initial rotational state, $ |JM \rangle$, of $^{16}$O$^{18}$C$^{32}$S. The time dependence of the
alignment cosine for the $|00 \rangle$, $|11 \rangle$, and $|22 \rangle$ states is shown in
\autoref{fig:nonadsim}(a)-(c)\footnote{Note that the calculated alignment cosine
   \protect{$\langle\cos^2\theta\rangle$} is different from the experimentally measured
   \protect{$\langle\cos^2\theta_{2D}\rangle$} but simulations have shown that their time-evolution
   quantitatively agrees.}. While the revival structures of the alignment cosines for the three
states are similar and in phase at the half- and full-period, the revival structures at the quarter-
and three-quarter-period are out of phase with respect to one another for states with
opposite parity, $(-1)^J$, i.e., the even-parity \ket{00} and \ket{22} states on the one hand and
the odd-parity \ket{11} state on the other. Panel (d) of \autoref{fig:nonadsim} shows the ensemble
average, $\langle\costhetasq\rangle=\Sigma_{J} w_J \Sigma_{M} \costhetasq_{JM} $, of the alignment
cosine \cite{friedrich-epjd:2006}, with $w_J$ the Boltzmann weights of the initial rotational states
(these are $w_0=0.567$; $w_1=0.396$; $w_2=0.036$ at $T=0.4$ K). The time dependence of
$\langle\costhetasq\rangle$ exhibits the same revival structure as the experimental dependence for
the undeflected beam, blue squares in \autoref{fig:nonadexp}. In particular, the quarter and
three-quarter period revivals are almost absent, due to the destructive interference of the
opposite-parity contributions to $\langle\costhetasq\rangle$ from the initial rotational states
\ket{00} and \ket{10}.

For the deflected beam (red circles in \autoref{fig:nonadexp}), the revival structure strikingly
differs from that of the undeflected beam: Firstly, a prominent quarter-period revival is now
present. As \autoref{fig:nonadsim}(a)-(c) indicate, this can only come about if most of the beam
molecules have the same parity. Since the position of the local minimum of the observed quarter
revival (at 20.9~ps) matches that due to the \ket{00} and the \ket{22} states (at 20.7~ps), but not
due to the \ket{11} state (at 19.7~ps), it must be the \ket{11} state whose concentration in the
beam has been diminished. The skewed structure with a local maximum (at 18.9~ps) followed by a local
minimum (at 20.9 ps) and then another (slightly lower) local maximum (at 21.7~ps) fits the simulated
\ket{00} trace well but is at odds with the simulated \ket{22} trace. Similarly, the shapes of the
observed half- and full-period revivals resemble closely those calculated for the \ket{00} but not
for the \ket{22} state. These observations corroborate what the simulated deflection curves,
\autoref{fig:deflection}, have suggested, namely that the molecular beam (at $y = 1.5$ mm) is
dominated by the \ket{00} state.

A quantitative assessment of the fraction of the \ket{11} state in the deflected beam could be
obtained by comparing the amplitude of the (left) local maximum at the quarter-period revival (at
18.9~ps) to the prompt alignment maximum (at 0.9~ps). This is because, as seen in
\autoref{fig:nonadsim}(b) and (c), molecules in the \ket{00} or \ket{11} states cause the local
maximum to rise or drop, respectively, in comparison with the prompt-alignment maximum. In this way,
we found that at least $92 \%$ of all the beam molecules must be in the \ket{00} state to account
for the observed revival amplitudes, \autoref{fig:nonadexp}. The deflection curves suggest $89 \%$
population of the \ket{00}, in fair agreement with the alignment revival data. We note that we do
not expect the simulated deflection curves to identify the $y$ position of the molecular profile to
an accuracy better than $0.1$ mm. This has repercussions for our ability to quantify the populations
of the states. For instance, at $y = 1.6$ mm the simulations yield a $94 \%$ population of the
\ket{00} state.

Our assessment of the populations of the rotational states in the beam was corroborated by yet
another piece of evidence: in the absence of the alignment pulse the measured alignment degree for
the undeflected beam is $\costhetasqtd = 0.50$, as it should for an isotropic ensemble, cf.\
$\costhetasqtd$ in \autoref{fig:nonadexp} just before the alignment pulse ($t = -1$~ps)). For the
deflected beam, the measured alignment degree has still the isotropic value of 0.50 or perhaps
marginally smaller. This is only possible if essentially only the isotropic ground state \ket{00}
with $\costhetasqtd =0.5$ is present in the deflected beam while the anisotropic states \ket{11} and
\ket{22} with $\costhetasqtd$ = 0.375 and 0.312, respectively, are nearly absent.

Secondly, the alignment trace in \autoref{fig:nonadexp} pertaining to the deflected beam exhibits a
significantly more pronounced modulation of the half- and full-period revivals than the trace for
the undeflected beam. Inspection of the half-period revival reveals that the local maximum of
$\costhetasqtd$, which is also the global maximum, is increased from 0.77 for the undeflected beam
to 0.84 for the deflected beam. We note that we did not measure the 3/4-period revival but,
according to our calculations, its local alignment maximum is well below that obtained at the
1/2-period revival. Likewise, the global minimum of $\costhetasqtd$ attained at the full-period
revival ($t = 81.6$~ps, see \autoref{fig:nonadexp}) decreases from 0.39 to 0.33. This means that not
only alignment but also anti-alignment, \ie confinement of the molecular axis to the plane
perpendicular to the polarization vector of the alignment pulse, is significantly enhanced by using
the deflected, state-selected beam compared with the direct beam.

The calculations displayed in \autoref{fig:nonadsim} predict that the alignment cosine,
$\costhetasq$, only increases from $0.86$ for a $0.4$ K thermal ensemble to $0.87$ for a pure
\ket{00} state. Even for a $1$ K beam, the alignment cosine is 0.85. Therefore, the much more
pronounced enhancement of the alignment observed experimentally is likely due to a rotational state
distribution in the direct beam which is not strictly Boltzmannian. As noted above, the measured
deflection curves, \autoref{fig:deflection}, indicate that the direct beam contains an excess of
molecules in the \ket{10} and possibly also the \ket{20} states. Our calculations show that the
alignment of these states at the half-period revival is significantly less than the alignment of the
\ket{00}, \ket{11} and \ket{22} states. As a result, an increased concentration of the \ket{10} and
the \ket{20} states in the molecular beam would lead to a weaker alignment than for a thermal beam
with a Boltzmann distribution of rotational states; such non-Boltzmann behavior of supersonic
expansions is a well-known phenomenon~\cite{Stolte:StateSelectedScattering:1988,
   Wu:JCP91:5278,Cappelletti:1999,Vattuone-Cappelletti-2010}. Overall, the deflector is well-suited
for generating molecular beams that can be particularly strongly aligned.

Selecting out rotational-ground-state molecules by electrostatic deflection will be particularly
effective for species with a small moment of inertia (large rotational constant), such as many
diatomic and small polyatomic molecules. Supersonic molecular beams of such species tend to have a
large population of the rotational ground state. The single-quantum-state selected beam may have
then an intensity of up to tens of percent of the undeflected beam. Examples include IBr, ICN, ClCN,
C$_2$HF, and CH$_3$I. For larger (and heavier) polyatomics, the number of states populated increases
rapidly with the moment of inertia and temperature, and thus selection of the rotational ground
state, if at all feasible, will occur at the expense of a much reduced beam intensity.

The ability to select out the absolute ground state of molecules by the deflection method has
implications for several research areas: (1) Laser-induced alignment will clearly benefit from a
state selection by the deflector, as demonstrated here. So will orientation based on pure optical
methods \cite{de_field-free_2009, oda_all-optical_2010} as well as on the combined electrostatic and
nonresonant radiative fields \cite{friedrich-herschbach:1999, haertelt-friedrich:2008,
   filsinger_quantum-state_2009, holmegaard_laser-induced_2009,
   ghafur_impulsive_2009,rouzee_optimization_2009}. In particular, related work relying on
state-selection by an electrostatic hexapole has already demonstrated that single-state molecular
beams are conducive to producing tightly aligned or oriented molecules
\cite{ghafur_impulsive_2009,rouzee_optimization_2009}. (2) Collision/reaction dynamics in crossed
molecular beams. (3) Ground-state molecules selected out by the deflector could be efficiently
optically decelerated and trapped \cite{Fulton:NP:2006}, thus opening an alternate route to trapping
for molecules which cannot be, for instance, Stark-decelerated.

\bibliography{jhn}

\end{document}